\documentclass{elsart}

\usepackage{amsmath, dsfont, graphicx}

\newcommand{\vek}[1]{\mathbf{#1}}
\newcommand{\uvek}[1]{\mathbf{\hat{#1}}}
\newcommand{\sfrac}[2]{\textstyle \frac{#1}{#2}}
\newcommand{\eps}{\varepsilon}

\renewcommand{\Re}{\text{Re\,}}

\newcommand{\order}{\mathcal{O}}
\newcommand{\setN}{\mathds{N}}
\newcommand{\setNz}{\mathds{N}_0}

\newcommand{\GSum}[4]{\sum_{\substack{#3=|#1-#2| \\ #1+#2+#3=2#4}}^{#1+#2}}
\newcommand{\GSumZ}[4]{\sum_{\substack{#3=|#1-#2| \\ #1+#2+#3=2#4,\, #4 \in
\mathds{Z}}}^{#1+#2}}
\newcommand{\FORM}{{\ttfamily FORM}}
\newcommand{\PARFORM}{{\ttfamily PARFORM}}
\newcommand{\SUMMER}{{\ttfamily SUMMER}}
\newcommand{\Mathematica}{{\ttfamily Mathematica}}
\newcommand{\Maple}{{\ttfamily Maple}}
\newcommand{\Fortran}{{\ttfamily FORTRAN}}

\newcommand{\MPI}{{\ttfamily MPI}}
\newcommand{\MPFUN}{{\ttfamily MPFUN}}
\newcommand{\type}[1]{{\ttfamily #1}}

\renewcommand{\thefootnote}{\fnsymbol{footnote}}

\begin{document}

\begin{frontmatter}
\begin{flushleft}
TTP05-08\\
SFB/CPP-05-16
\end{flushleft}

\title{Calculation of Massless Feynman Integrals using Harmonic Sums}

\author{Stefan Bekavac}

\address{Institut f\"ur Theoretische Teilchenphysik, Universit\"at Karlsruhe (TH)}

\begin{abstract}
A method for the evaluation of the $\varepsilon$-expansion of multi-loop
massless Feynman integrals is introduced.
This method is based on the Gegenbauer polynomial technique and the expansion of
the Gamma function in terms of harmonic sums. Algorithms for the evaluation of nested and
harmonic sums are used to reduce the expressions to get analytical or numerical
results for the expansion coefficients. Methods to increase the precision of
numerical results are discussed. 
\end{abstract}


\end{frontmatter}


\section{Introduction}
\label{introduction}
One of the most important aims of particle physics is the test of models, particularly the standard model.
Testing the standard model means,
on one hand, to increase the accuracy of its parameters and, on the other hand,
the search for deviations of its predictions from measured quantities to
discover new effects. 
 To do
this experiments are accomplished with increasing accuracy of measurement.
Therefore theoretical calculations have to be performed with increasing
numerical precision. 

A parameter of the standard model of particular interest is the strong coupling constant $\alpha_s$. It
can, for example, be determined via the so-called $R$-ratio, the normalized total cross
section of the process $e^+\,e^- \to \text{hadrons}$,
\begin{equation}
R(s) = \frac{\sigma({e^+\,e^- \to \text{hadrons}})}{\sigma(e^+\,e^- \to \mu^+\,\mu^-)}.
\end{equation}

In order to achieve high precision one has, within perturbative quantum field theory, to calculate high orders 
of the perturbation series. This means, we need to calculate large numbers of more
and more complicated Feynman diagrams.
Therefore different methods for the evaluation of such multiloop Feynman diagrams 
have been developed. (See e.g. \cite{Harlander,Grozin,Smirnov:Feynman}.)

The quantity $R(s)$ is nowadays known to $\order(\alpha_s^3)$ precision
\cite{Gorishnii,Chetyrkin:Rs} and first
results in $\order(\alpha_s^4)$ have been published \cite{Baikov:epem}.
For a review see Refs. \cite{Chetyrkin:Rep,Rhad}.

A prescription to solve problems with a large number of complicated Feynman diagrams is to reduce them to a set of
master integrals, e.g. by means of the traditional integration by parts method, and then evaluate these. 

The aim of the project described in this article is to develop a method for the
calculation of massless integrals. More precisely, we want to consider the  
planar ($P$) and non-planar ($N$) three-loop propagator diagrams:

\hfill
$P = $ \parbox{4cm}{\includegraphics[height=2.5cm]{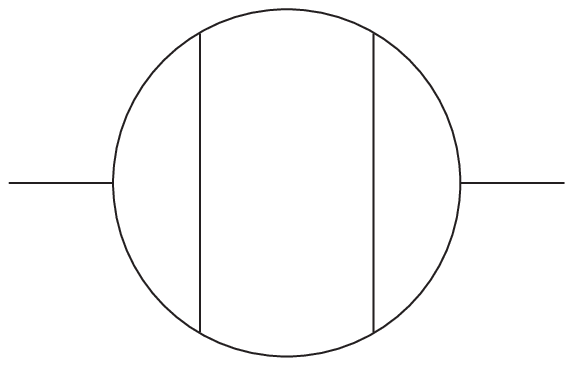}}
\hfill
$N = $ \parbox{4cm}{\includegraphics[height=2.5cm]{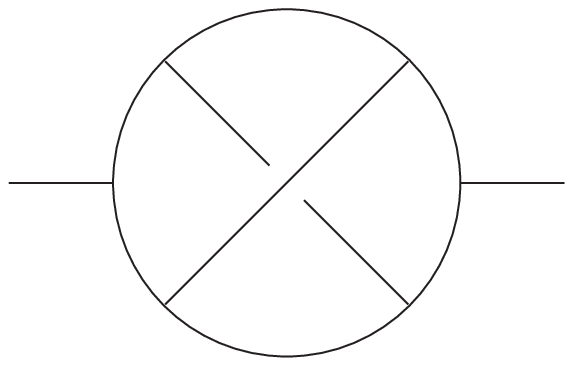}}
\hfill{}

They can be considered as the master integrals of the corresponding class of Feynman
diagrams, which are ``building blocks'' of the calculations for $R(s)$ in
$\order(\alpha_s^4)$.  \\

We are interested in the $\eps$-expansion of the diagrams, where $\eps$ is the
parameter of {\em dimensional regularization}. This means, the integrals are
formally calculated in $D=4-2\eps$ dimensions and then expanded in $\eps$. By
means of this procedure UV divergencies of diagrams manifest themselves as poles
in $\eps$ and can then be cancelled by means of {\em renormalization}.

It is thus essential for renormalization to know the pole structure of a Feynman
diagram and the finite part of the expansion. However, when one has to deal with
products of graphs with poles as in the case of the following diagram,

\hfill \includegraphics[height=2cm]{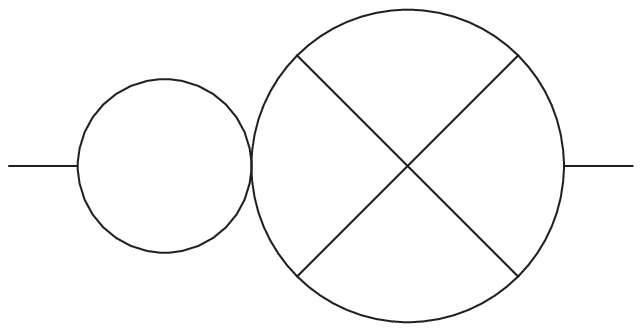} \hfill{}

which appears within the mentioned calculations of $R(s)$,
one has to calculate higher $\eps$ orders of the second
factor in order to ensure, that the result is correct up to the finite term.  
This is why we are interested in higher orders of the $\eps$ expansion of the
diagrams under consideration. 

The $\order(\eps^1)$ terms of the diagrams $N$ and $P$ are necessary for the
computation of the $\beta$ function of the scalar $\phi^4$ model at five-loop
level. They have been known since long from Ref. \cite{Gorishnii:phi4} and
\cite{Kazakov:MOU}.

The calculation of the planar diagram is somewhat simpler than that of the
nonplanar one, because it could be related to the generic massless two-loop
two-point function by means of integration by parts \cite{Chetyrkin:IBP}.
The latter was studied in many works. e.g.
\cite{Kazakov:MOU,Chetyrkin,Kazakov:uni,Broadhurst:2loop,Barfoot,Gorishnii:approach}
and references therein. Recently, in Ref. \cite{Bierenbaum} it was in a sense
computed analytically.
Thus one can say that the planar diagram is known to arbitrary order and
certainly to sufficient precision for practical needs. 

The known results for the diagrams $P$ and $N$ read \cite{Kazakov:MOU,Chetyrkin:pers}
\begin{equation}
N=\frac{1}{(4\pi)^6} G^3(\eps) (k^2)^{-2-3\eps} \tilde{N}_S,
\end{equation} 
\begin{equation}
\label{Ergebnis_N2}
\tilde{N}_S =
20\, \zeta_5+\eps\,(68\, \zeta_3^2-80\, \zeta_5+50\, \zeta_6)+\mathcal{O}(\eps^2)
\end{equation}
and
\begin{equation}
\label{Ergebnis_P}
P=\frac{1}{(4\pi)^6}G^3(\eps)(k^2)^{-2-3\eps} \tilde{P}_S,
\end{equation}
\begin{multline}
\tilde{P}_S=20\zeta_5 + \eps \,(44\zeta_3^2 -80\zeta_5+50\zeta_6)\\
+\eps^2
\left(-176\zeta_3^2+132\zeta_3\,\zeta_4+80\zeta_5-200\zeta_6+317\zeta_7\right) +\mathcal{O}(\eps^3).
\end{multline}

In the following we denote the expansion coefficients of order $n$
by $|_{\eps^n}$, e.g. $\tilde{N}_S|_{\eps^1} = 68\, \zeta_3^2-80\, \zeta_5+50\,
\zeta_6$.

These numbers have been used to check the method and to examine the properties
of the numerical algorithms described later.

The determination of the $\eps^2$ coefficient of $N$
is one of the aims of the project presented in this paper. An independent
evaluation can be found in Ref. \cite{Baikov:unpublished}, the result was given
in Ref. \cite{Chetyrkin:Talk}. A numerical approach to the calculation of
integrals of this type is given e.g. in Ref. \cite{Binoth}.

The paper is organized as follows:
Section \ref{GPXT} shows how a representation of the diagrams $N$ and $P$ in
terms of sums can be found
by means of the Gegenbauer polynomial method. 
Section \ref{Sums} gives an overview 
over the Harmonic Sums and algorithms to deal with them. 
It is shown how the summation algorithms are applied to evaluate the sums
representing the diagrams.
For the $N$ diagram an analytical solution by means of the described
algorithms is not yet possible. In this case one can get numerical results by summing up the series.
Section \ref{convergence} reviews methods to accelerate the convergence of such
processes to get better results. 
By means of these numerical methods it is possible to calculate the expansion
coefficients with high precision. For the non-planar diagram this is described in Section \ref{nonplanar}.
The planar topology can be calculated analytically with the help of the summation
algorithms, which is demonstrated in Section \ref{planar}.


\section{The Gegenbauer Polynomial x-Space Technique}
\label{GPXT}
\begin{figure}[h]
\begin{center}
\includegraphics[height=5cm]{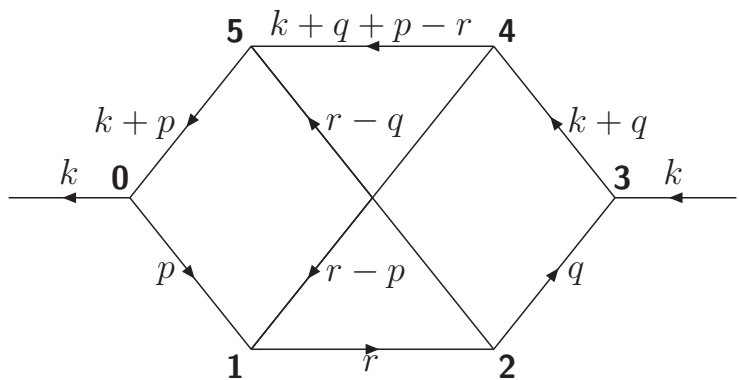}
\caption{Labeling of the momenta and vertices for diagram $N$ \label{N_labels}}  
\end{center}
\end{figure}

The Gegenbauer Polynomial x-Space Technique (GPXT) to evaluate Feynman Diagrams in coordinate
space was introduced in Ref. \cite{Chetyrkin}. Applications can be found e.g. in
Refs. \cite{Celmaster,Broadhurst:Knots}, an extension of the technique in Ref.
\cite{Kotikov}.
In this
Section we will apply the method to the diagram $N$. The results for the planar
diagram can be obtained analogously. Labeling the momenta
as shown in Fig. \ref{N_labels} the integral
representation of $N$ reads

\begin{multline}
N = \frac{1}{(2\pi)^{3D}} \int d^Dp\, d^Dq\, d^Dr\, \\
 \frac{1}{p^2\, r^2\, q^2\,
(k+q)^2\, (k+q+p-r)^2\, (k+p)^2\, (r-q)^2\, (r-p)^2}.
\end{multline}

By means of Fourier transformation
\begin{equation}
\frac{1}{k^2} = \frac{\Gamma(\lambda)}{\pi^{\lambda+1}}\, \int
\frac{e^{2i\vek{k}\vek{x}}d^Dx}{(x^2)^\lambda},
\end{equation}

where $\lambda = 1-\eps = \frac{D}{2}-1$,
the integrand is transformed to position space. The integrals over the
momenta can then be evaluated to delta functions:
\begin{equation}
\int e^{2i\vek{p}\vek{x}}\, d^Dp = \pi^D\, \delta(x).
\end{equation}

After the evaluation of the delta functions and choosing an appropriate
coordinate system (see Fig. \ref{N_labels}, where the root vertex is denoted by {\sffamily 0})
the denominators get a simple and uniform shape:

\begin{equation}
\begin{split}
N &= 
\frac{\Gamma^8(\lambda)}{2^{6(\lambda+1)}\, \pi^{8(\lambda+1)}}
\int d^Dx_1\, d^Dx_2\, d^Dx_3\, d^Dx_4\, d^Dx_5 \; e^{-2i\vek{k}\vek{x}_3}
\frac{1}
{x_1^{2\lambda}\, x_5^{2\lambda}}\\
&\times \frac{1}{(x_2-x_1)^{2\lambda}\, (x_3-x_2)^{2\lambda}\,
(x_4-x_3)^{2\lambda}\, (x_5-x_4)^{2\lambda}\, (x_1-x_4)^{2\lambda}\, (x_5-x_2)^{2\lambda}\,
}
\end{split}
\end{equation}

Spherical coordinates are introduced by means of
\begin{equation}
d^Dx = \frac{1}{2}\, \frac{2\pi^{\lambda+1}}{\Gamma(\lambda+1)}\,
r^\lambda\,dr\,d\uvek{x},
\end{equation}
where
\begin{equation*}
r=x^2 \quad \text{and} \quad \uvek{x} = \frac{x}{\sqrt{r}} .
\end{equation*}
Now the propagator terms and the exponential function can be
expanded in Gegenbauer polynomials using formulae (\ref{GGP-Entwicklung}) and
(\ref{Exp-Entwicklung}) of appendix \ref{Anhang_GP}. This leads to a representation of $N$ as a multiple
sum

\begin{eqnarray}
\label{7fachSumme}
N &=& \frac{\Gamma^3(\lambda)}{\lambda^5\,(4\pi)^{3(\lambda+1)}} 
     \sum_j \sum_l \sum_m \sum_n \sum_s \sum_t \sum_u \\
  &\times&  
          \int dr_1 d\uvek{x}_1 \int dr_2 d\uvek{x}_2 \int dr_3 d\uvek{x}_3 \int
	  dr_4 d\uvek{x}_4 \int dr_5 d\uvek{x}_5 \nonumber \\
  &\times& \Gamma(\lambda)\, i^j\,(j+\lambda)\, C_j^\lambda(\uvek{k} \cdot
  \uvek{x}_3)\, (k^2r_3)^{j/2}\, j_{j+\lambda}(k^2r_3) \nonumber \\
  &\times&
  \frac{r_1^\lambda\,r_2^\lambda\,r_3^\lambda\,r_4^\lambda\,r_5^\lambda}
   {r_1^\lambda\,r_5^\lambda} \;  C_l^\lambda( \uvek{x}_1 \cdot \uvek{x}_2 ) \,
  C_u^\lambda( \uvek{x}_2 \cdot \uvek{x}_3 ) \,  \nonumber \\
  &\times&
  C_n^\lambda( \uvek{x}_3 \cdot \uvek{x}_4 ) \,
  C_s^\lambda( \uvek{x}_4 \cdot \uvek{x}_5 ) \,
  C_t^\lambda( \uvek{x}_1 \cdot \uvek{x}_4 ) \,
 C_m^\lambda( \uvek{x}_2 \cdot \uvek{x}_5 ) \nonumber\\
      &\times& M_\lambda^{l}(r_1,r_2) \, M_\lambda^{n}(r_2,r_3) \,
            M_\lambda^{n}(r_3,r_4) \, M_\lambda^{m}(r_4,r_5) \,
	    M_\lambda^{l}(r_1,r_4) \, M_\lambda^{m}(r_2,r_5) \nonumber ,
\end{eqnarray}

where 
\begin{equation}
M_\lambda^n(r_i,r_j) = \frac{1}{\max(r_i,r_j)^\lambda} \, \left< \frac{r_i}{r_j} \right>^{n/2}
= \frac{1}{\max(r_i,r_j)^\lambda} \, \min \left( \frac{r_i}{r_j} , \frac{r_j}{r_i} \right)^{n/2}.
\label{M_def}
\end{equation}

The angular integrations can be evaluated using the properties of the
Gegenbauer polynomials.
Each application of the orthogonality relation (\ref{GGP_ORS}) reduces two
$C^\lambda_n$ to one and produces a Kronecker delta. Finally, one arrives at the
expression

\begin{eqnarray}
\label{Radialintegral}
N &=& \frac{\Gamma^4(\lambda)}{(4\pi)^{3(\lambda+1)}} 
     \sum_{l=0}^\infty \sum_{m=0}^\infty \GSumZ{l}{m}{n}{g}
     \int dr_1 \int dr_2 \int dr_3 \int dr_4 \int dr_5\\
  &\times&  
    \frac{\Gamma(n+2\lambda)}{n!\, \Gamma(2\lambda)}\,
     \frac{D_\lambda(l,m;n)}{(l+\lambda)\,(m+\lambda)\,(n+\lambda)^2}
     \; j_\lambda(k^2r_3)	  
\frac{r_1^\lambda\;r_2^\lambda\;r_3^\lambda\;r_4^\lambda\;r_5^\lambda}{r_1^\lambda\;r_5^\lambda} \nonumber \\
   &\times& M_\lambda^{l}(r_1,r_2) \, M_\lambda^{n}(r_2,r_3) \,
            M_\lambda^{n}(r_3,r_4) \, M_\lambda^{m}(r_4,r_5) \,
	    M_\lambda^{l}(r_1,r_4) \, M_\lambda^{m}(r_2,r_5) \nonumber .
\end{eqnarray}

The complicated structure of the triple sum is introduced by Eq.
(\ref{CC-Summe}), which is applied
when there are two polynomials with the same arguments.

To handle the radial integrals the region of integration is distributed in 120
regions according to the relative size of $r_1,r_2,r_3,r_4,r_5$. In each region the
expressions $M_\lambda^n$, Eq. (\ref{M_def}), evaluate
to rational expressions. For simplification $k^2$ is set to 1; the $k$-dependence
can later be restored by dimensional arguments.
Thus we are lead to integrals like

\begin{equation*}
\int_0^{r_3} dr_2\, r_2^\frac{m+n-l}{2} \, \int_0^{r_2} dr_1\, r_1^l\,
    \int_0^\infty dr_3\, j_\lambda(r_3) \, \int_{r_4}^\infty dr_5\,
    r_5^{-m-2\lambda}\, \int_{r_3}^\infty dr_4 \, r_4^\frac{m-n-l-2\lambda}{2}
\end{equation*}
\begin{equation*}
=\frac{4}{(l+1)\,(l+m+n+4)\,(m+2\lambda-1)\,(l+m+n+6\lambda-4)}
\frac{\Gamma(5-3\lambda)}{\Gamma(4\lambda-4)} .
\label{radial}
\end{equation*}

Actually, due to the symmetry of the diagram, only 30 of these 120 terms are
different. The calculation of these can be automated.
\footnote{We thank K. Chetyrkin for his \Mathematica{} implementation of this
procedure.}

After performing all integrations and rewriting $\lambda=1-\eps$ the integral $N$ is expressed 
in terms of a triple sum:
\begin{equation}
N = C \cdot N_S,
\end{equation}
\begin{multline}
\label{NS}
N_S=\sum_{l=0}^\infty \sum_{m=0}^\infty \GSum{l}{m}{n}{g} R(\eps,l,m,n)\,
\frac{(1-\varepsilon)^4}{\Gamma(2-2\varepsilon)\,\Gamma^2(1-\varepsilon)}\\
\times \frac{1}{(l+1-\varepsilon)\,(m+1-\varepsilon)\,(n+1-\varepsilon)}\,
\frac{\Gamma(g+2-2\varepsilon)}{\Gamma(g+2-\varepsilon)}\\
\times \frac{\Gamma(g-l+1-\varepsilon)\,\Gamma(g-m+1-\varepsilon)\,\Gamma(g-n+1-\varepsilon)}
{\Gamma(g-l+1)\, \Gamma(g-m+1)\, \Gamma(g-n+1)},
\end{multline}

where
{\footnotesize
\begin{equation*}
C = \frac{\Gamma^4(1-\eps)\, \Gamma(2+3\eps)}{(4\,\pi)^{3(2-\eps)}\,(1-\eps)^4\,
\Gamma(-4\,\eps)},
\end{equation*}

\begin{multline*}
R(\eps,l,m,n)=\frac{4}{(l+1)\,(l+m+n+4)\,(m-2\eps+1)\,(l+m+n-6\eps+2)}\\
+\frac{8}{(l+1)\,(l+m+n+4)\,(l+m+n-4\eps+2)\,(l+m+n-6\eps+2)}\\
+ \ldots \text{(118 similar terms)}.
\end{multline*}
}

Finally, according to Ref. \cite{Chetyrkin}, this result can be cast into the
so-called $G$-Form. Rewriting
\begin{equation}
C = \frac{1}{(4\pi)^6}\,G^3(\eps)\cdot(-4\eps-4\eps^2+32\eps^3+\mathcal{O}(\eps^4))
\end{equation}

one obtains
\begin{equation}
N=\frac{1}{(4\pi)^6} G^3(\eps) (k^2)^{-2-3\eps} \tilde{N}_S,
\end{equation}
\begin{equation}
\label{nstilde}
\tilde{N}_S = (-4\,\eps-4\,\eps^2+32\,\eps^3+\mathcal{O}(\eps^4)) N_S.
\end{equation}

$N_S$ can now be expanded in powers of $\eps$. For each
order in $\eps$ a triple sum has to be evaluated. The calculation of these sums is the aim of the rest of the
paper.


\section{Summation Algorithms}
\label{Sums}
In this Section the Harmonic Sums are introduced and their relationship to the
problem under consideration is shown. 

\subsection{Harmonic Sums}

Harmonic Sums are defined as 
\cite{Vermaseren,Uwer,Concrete}
\begin{equation}
S_m(n) = \sum_{i=1}^n \frac{1}{i^m},
\end{equation}
which, for $n=\infty$, leads to  Riemann's zeta function:
\begin{equation}
S_m(\infty) = \zeta(m) \qquad  (m \geq 2).
\end{equation}
Higher Harmonic Sums are defined
\cite{Vermaseren} recursively via the relation
\begin{equation}
S_{m,j_1,...,j_p}(n)=\sum_{i=1}^n \frac{1}{i^m} S_{j_1,...,j_p}(i).
\end{equation}

The number of indices of a harmonic sum is called its {\em depth}, the sum of
the absolute values of the indices its {\em weight}.

A negative index denotes an alternating sum,
\footnote{ 
Note that this is the convention used in \cite{Vermaseren}.
Other authors denote a
factor $(-1)^m$ by overlining the index. In this case a negative sign of the
index indicates --- consequently --- a positive power of the corresponding variable.}

\begin{equation}
S_{-m}(n) = \sum_{i=1}^n \frac{(-1)^i}{i^m}.
\end{equation}

In Ref. \cite{Vermaseren} different algorithms are described to deal
with expressions of the form
\begin{equation}
S_{a,b,c}(n) = \sum_{i=1}^n \frac{1}{i^a} \sum_{j=1}^i \frac{1}{j^b}
\sum_{k=1}^j \frac{1}{k^c}
\label{sabc}
\end{equation}
and sums over such. These are implemented in the program package \SUMMER\
written in
\FORM\ \cite{Vermaseren:NewForm} which is available under the URL \\
\type{http://www.nikhef.nl/\~{}form/FORMdistribution/packages/summer}.

The harmonic sums are related to the psi (digamma) function, which is the logarithmic
derivative of the Gamma function, and its derivatives, the polygamma functions
$\psi^{(n)}(x)$,
\begin{equation}
\psi(x) = \psi^{(0)}(x)= \frac{d}{dx} \ln \Gamma(x) = \frac{\Gamma^\prime
(x)}{\Gamma(x)},
\end{equation}
\begin{equation}
\psi^{(n)}(x) = \frac{d^n}{dx^n} \psi(x),
\end{equation}
via the following relations \cite{Prudnikov}:
\begin{equation}
\psi^{(0)}(n) = -\gamma + \sum_{k=1}^{n-1} \frac{1}{k} = -\gamma + S_1(n-1),
\label{psi0->S}
\end{equation}
\begin{equation}
\psi^{(m)}(n) = (-1)^m m! \left[ -\zeta(m+1) + S_{m+1}(n-1) \right].
\label{psin->S}
\end{equation}

Using these relations the Gamma function can be expanded in terms of harmonic
sums (cf. Ref. \cite{Veretin})
\begin{equation}
\frac{\Gamma(n+\varepsilon)}{\Gamma(1+\varepsilon)} = \Gamma(n) \,\exp 
\left\{ \sum_{m=1}^\infty \eps^m \,\sfrac{(-1)^{m-1}}{m} \, S_m(n-1) \right\}.
\label{Gamma-Entwicklung}
\end{equation}

The expansion of the exponential function leads to the expansion formulae used in
the following:
\begin{eqnarray}
\label{Gamma-S-Entwicklung}
\frac{\Gamma(n+\eps)}{\Gamma(1+\eps)\,\Gamma(n)} &=& 1 + \eps S_1(n-1) \\
&&+ \sfrac{1}{2}\, \eps^2 \left[ S_1^2(n-1) - S_2(n-1) \right] \nonumber \\
&&+ \sfrac{1}{6}\, \eps^3 \left[ S_1^3(n-1) - 3 S_1(n-1) S_2(n-1) + 2 S_3(n-1) \right] \nonumber \\
&& + \order(\eps^4), \nonumber \\
\label{1/Gamma-S-Entwicklung}
\frac{\Gamma(1+\eps)\,\Gamma(n)}{\Gamma(n+\eps)} &=& 1 - \eps S_1(n-1) \\
&&+ \sfrac{1}{2}\, \eps^2 \left[ S_1^2(n-1) + S_2(n-1) \right] \nonumber \\
&&- \sfrac{1}{6}\, \eps^3 \left[ S_1^3(n-1) + 3 S_1(n-1) S_2(n-1) + 2 S_3(n-1) \right] \nonumber \\
&&+ \order(\eps^4). \nonumber
\end{eqnarray}

\subsection{Simplification of the sums}
The structure of the sum given in Eq. (\ref{NS}),
\begin{equation}
N = \sum_{l=0}^\infty \sum_{m=0}^\infty
\sum_{\substack{n=|l-m|\\l+m+n=2g}}^{l+m} T(l,m,n), \quad g \in \mathds{N},
\end{equation}
 particularly the boundaries of the innermost
sum, is very difficult to handle with symbolic manipulation programs. Therefore
the first step is the simplification of this sum.

\begin{figure}[h]
\begin{center}
\includegraphics{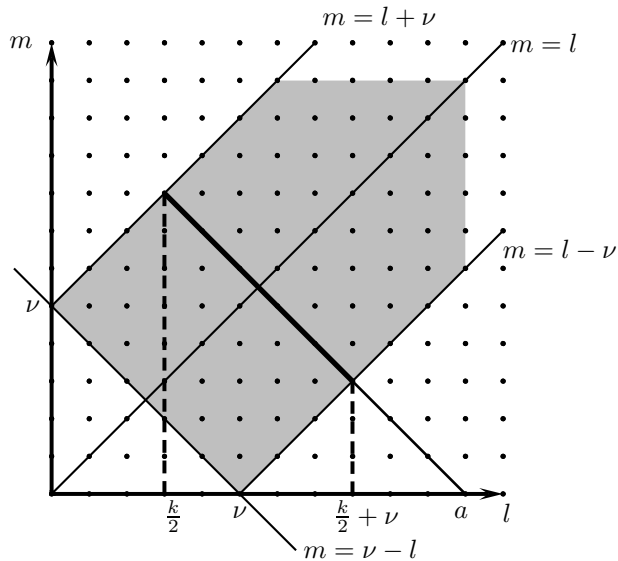}
\caption{Area of summation for fixed $n=\nu$. \label{Summationsdiagramm}}
\end{center}
\end{figure}

For a given $n=\nu$ we have to sum over pairs of indices $(l,m)$, which satisfy
\begin{equation}
|l-m| \leq \nu \leq l+m .
\end{equation}
Neglecting the condition $l+m+n=2g$ at first,
this is equivalent to
\begin{equation}
m \geq l-\nu \;\wedge\; m \leq l+\nu \;\wedge\; m \geq \nu-l,
\label{Ungleichunglm}
\end{equation}
which is described by the rectangular area in the $(l,m)$-plane shown in
Fig. \ref{Summationsdiagramm}.

Taking into account the condition $l+m+n=2g$ this means, that the sum ranges over all 
combinations of indices in this marked area
whose sum is even.

Let us in a next step consider pairs $(l,m)$ with constant sum $a=l+m$, i.e. points on straight
lines parallel to the second bisector.
The sum starts with the line $m=\nu-l$ ($a=\nu$) and reaches over the shaded
rectangle in Fig. \ref{Summationsdiagramm} to infinity. Taking into account the 
symmetry of the terms it is sufficient to restrict oneself to the area below or
over the first bisector.
The contributing points on such a line can be described by the condition $k \leq
2l \leq k+2\nu$, where $\nu+k=a$
(cf. Fig. \ref{Summationsdiagramm}).

If we parametrize $k=2\kappa$ and $k=2\kappa+1$, the addends
read $F(l,\nu+2\kappa-l,\nu)$ and $F(l,\nu+2\kappa+1-l,\nu)$, respectively, and 
it is obvious, that in the second case the sum of the arguments
can never be even, while
the terms of the first kind automatically fulfill the condition.

Therefore the original sum is transformed to a much simpler form:
\begin{equation}
\sum_{n=1}^\infty \sum_{\kappa=0}^\infty \sum_{l=\kappa}^{n+\kappa}
F(l,n+2\kappa-l,n)
= \sum_{n=1}^\infty \sum_{\kappa=0}^\infty \sum_{l=0}^{n}
F(l+\kappa,n+\kappa-l,n).
\label{FormSumme}
\end{equation}

If the expansions (\ref{Gamma-S-Entwicklung},\ref{1/Gamma-S-Entwicklung}) are applied to the sum
in Eq. (\ref{7fachSumme}), it is
represented as a triple sum over rational expressions and products of such and
harmonic sums. In the following some algorithms are discussed to simplify such
objects.

\subsection{Rational Expressions}

To sum expressions of the form
\begin{equation}
S=\sum_{x=\alpha}^\beta
\frac{P(x)}{(x+a_1)\,(x+a_2)\,\ldots(x+b_1)^2\,(x+b_2)^2\,\ldots\,(x+q_1)^n\,(x+q_2)^n\,\ldots},
\label{SummeGRF}
\end{equation}
where $P(x)$ is a polynomial in $x$, one first performs partional fractioning. This leads to simpler summands in which $x$ appears only once in the
denominator. The resulting sums can be evaluated easily in terms of zeta or psi
functions by means of the formula \cite{Summation}

\begin{multline}
\sum_{x=\alpha}^\beta \frac{1}{(ax+b)^n} = \frac{1}{a^n} \left(
\zeta(n,\frac{b}{a}+\alpha) - \zeta(n,\frac{b}{a}+\beta+1)
\right)  \\
= \frac{(-1)^n}{(n-1)!}\, \frac{1}{a^n} \,\left(
\psi^{(n-1)}(\frac{b}{a}+\alpha )
- \psi^{(n-1)}(\frac{b}{a}+\beta+1) \right).
\label{PsiSummeallg}
\end{multline}

If the upper limit is infinity, the second psi function is zero for $n>1$.

\subsection{Algorithms for Nested Sums}
In Ref. \cite{Uwer} different algorithms have been published to evaluate nested
sums using the properties of higher harmonic sums and more general structures
like the so-called $S$- and $Z$-sums. 
For our purpose some special cases of the algorithms A and B of this work are needed.

If an addend is not of the simple form $S(a)/a^m$, but contains an offset in the
denominator, which must not be symbolic, the summand can be reduced recursively

\begin{equation}
\sum_{i=1}^n \frac{S_{kl}(i)}{(i+c)^m}  = \sum_{i=1}^n \frac{S_{kl}(i)}{(i+c-1)^m}
 - \sum_{i=1}^n \frac{1}{(i+c-1)^m} \frac{S_l(i)}{i^k}  +
\frac{S_{kl}(n)}{(n+c)^m}.
\label{AlgorithmusA}
\end{equation}
This is a special case of Algorithm A. Of course, this is also valid for sums of
depth greater than 2. 

Algorithm B describes the recursive simplification of sums where products of
harmonic sums appear. 

An expression of the form
\begin{equation}
\sum_{a=1}^\infty \sum_{b=1}^a \frac{S_k(b) \, S_l(a-b)}{a^m \, b^n}
\label{AlgBTerm}
\end{equation}
can be rewritten by inserting the definition of $S_l$. Then shifting
variables, reordering of summations and partial fractioning yields to
expressions which can be evaluated immediately and such of the same form as the
original expression to which the procedure is applied again. Thus the sum is
reduced recursively to simpler expressions and finally evaluated.

\subsection{Products of Harmonic Sums}
In Ref. \cite{Bluemlein2000} algebraic relations
between harmonic sums have been studied.
For harmonic sums with identical arguments the permutation relation holds:

\begin{equation}
S_{m,n} + S_{n,m} = S_m S_n + S_{m \wedge n},
\end{equation}
where
\begin{equation}
m \wedge n := \text{sign}(m)\, \text{sign}(n)\, (|m|+|n|).
\end{equation}

This allows us to express the product of two harmonic sums in the following way:
\begin{equation}
S_m(x) S_n(x) = S_{m,n}(x) + S_{n,m}(x) - S_{m \wedge n}(x)
\label{SS->S+S}
\end{equation}

For products with higher harmonic sums one has relations like 
\begin{equation}
S_a(x) \cdot S_{bc}(x) = S_{abc}(x) + S_{bac}(x) + S_{bca}(x) - S_{a \wedge
b,c}(x) - S_{b,a \wedge c}(x).
\end{equation}
More relations of this kind can be found in \cite{Bluemlein2003}.

\subsection{Harmonic Sums with Double Argument}
In the following also sums with double argument $S_k(2m)$ will appear. These cannot be
handled with the algorithms described up to now. 
In Ref. \cite{Weinzierl} transformations  of such objects to simpler sums have
been discussed.
From this work we need the relation 
\begin{equation}
\sum_{i=1}^\infty \frac{1}{(i-x)^m} S_k(2i) = 2^{m-1} \sum_{i=1}^\infty
\frac{1}{i^m} S_k(i+2x) + 2^{m-1} \sum_{i=1}^\infty \frac{(-1)^i}{i^m} S_k(i+2x)
\label{S2x}
\end{equation}
which relates the sum over $S(2k)$ to two sums over $S(k)$, one of which is
alternating. 
The opposite operation, the transformation of a sum with simple argument to one
with double argument, can be done by means of a special case of the refinement formula
\cite{Weinzierl}
\begin{equation}
S_m(n) = 2^{m-1} \left( S_m(2n) + S_{-m}(2n) \right) .
\label{S(n)->S(2n)}
\end{equation}

Generally, we can write a sum of the form
\begin{equation}
\sum_{i=1}^\infty f(2i)
\end{equation}
in the following way:
\begin{equation}
\sum_{i \in \setN} f(2i) = \sum_{i\, \text{even}} f(i) =  \frac{1}{2} \left[ \sum_{i \in \setN} f(i) -
\sum_{i \in \setN} (-1)^i f(i) \right].
\label{sumf(2x)}
\end{equation}
The sum over all even numbers is replaced by the sum over all integers minus the
sum over the odd numbers. This is achieved by
subtracting an alternating sum and the division by 2. 

\subsection{Multiple Infinite Sums}
\label{nestedsums}
Finally, we need some formulae to convert nested sums to multiple infinite sums
or vice versa. Two important relations are given by

\begin{equation}
\sum_{l=0}^\infty \sum_{m=0}^l \phi(l,m) = \sum_{m=0}^\infty \sum_{l=0}^\infty
\phi(l+m,m),
\label{sum-l0i-m0l}
\end{equation}

\begin{multline}
\sum_{l=1}^\infty \sum_{m=1}^{2l} \phi(l,m) =\\
\sum_{m=0}^\infty \sum_{l=0}^\infty \phi(l+m+1,2m+1) + 
\sum_{m=0}^\infty \sum_{l=0}^\infty \phi(l+m+1,2m+2).
\label{sum2l}
\end{multline}

\begin{figure}[h]
\begin{center}
\includegraphics{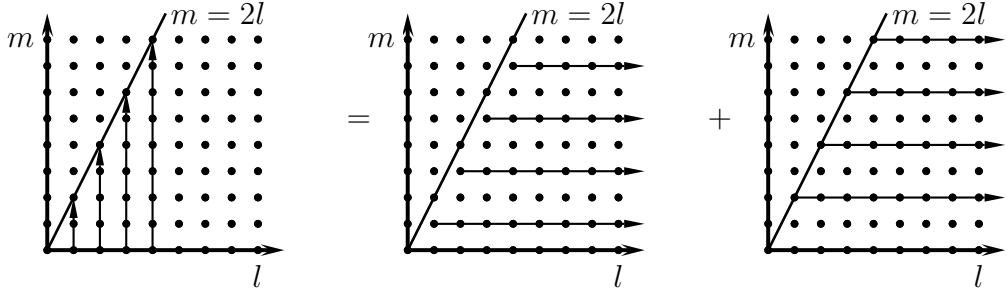}
\caption{Splitting of a sum of type (\ref{sum2l}) in two sums \label{sum2l_1}}
\end{center}
\end{figure}

Eq. (\ref{sum2l}) can be obtained by splitting the sum on the left-hand side into
two sums, which is visualized in Fig. \ref{sum2l_1}, and paramaterizing the summation variable $m=2\mu$ and $m=2\mu+1$,
respectively. This leads to sums of the type $\sum_{m=0}^\infty
\sum_{n=m}^\infty$. These can be converted to a double infinite series by
shifting the inner summation variable.

If the argument of the appearing harmonic sums is a sum of two or more variables,
the following relations are helpful:
\begin{equation}
\sum_{l=0}^\infty \sum_{m=0}^\infty \phi(l,m) \, S(l+m)
= \sum_{n=0}^\infty \sum_{l=0}^n \phi(l,n-l) \, S(n),
\end{equation}

\begin{multline}
\sum_{l=0}^\infty \sum_{m=0}^\infty \phi(l,m)\,S(l+2m)\\
= \sum_{n=0}^\infty \sum_{m=0}^n \phi(2n-2m,m)\,S(2n) + \sum_{n=0}^\infty
\sum_{m=0}^n \phi(2n-2m+1,m)\,S(2n+1).
\end{multline}


\section{Convergence Acceleration}
\label{convergence}
If the sums cannot be completely simplified analytically, one can at least get a
numerical result by summing up the terms with the computer. However, in our case
the sum
converges very slowly. In this Section we briefly sketch the method of {\em nonlinear
sequence transformations}, by means of which one can improve the convergence of a series. 
For further information we refer to Ref. \cite{Weniger} and references cited therein.

We consider a sequence $\{a_n\}$ and the corresponding infinite series 
\begin{equation}
S=\sum_{k=0}^\infty a_k
\end{equation}
with the partial sums
\begin{equation}
S_n = \sum_{k=0}^n a_k,
\end{equation}
which themselves form a sequence $\{S_n\}$ that converges to the value S. 
The question is now, whether a different sequence $\{S_n^\prime\}$ (or
$\{a_n^\prime\}$)
exists that converges to the same value $S$ but has a better convergence
behaviour. 


The sequence $\{S_n^\prime\}$ is said to {\em converge faster} than $\{S_n\}$
to the common limit  $S$, if
\begin{equation}
\lim_{n \to \infty} \frac{S_n^\prime-S}{S_n-S} = 0.
\end{equation}
In practice this means that $S_N^\prime$ is a better approximation to the
exact value of the limit than $S_N$, if the partial sum of order $N$ is known. 

Convergence acceleration methods can be found with the help of model sequences.
This means, one investigates a particular class of sequences and constructs a
transformation, i.e. a prescription how to calculate the elements of a new sequence
$\{S_n^\prime\}$  out of the elements of $\{S_n\}$, which is then exactly valid 
for the considered sequences. The prescriptions
one has found in this way are often able to improve the convergence of different
sequences of similar structure. 

One of the oldest and best known methods is {\em Aitken's
$\Delta^2$-process}. The partial sums are transformed according to the rule
\begin{equation}
S_n^\prime = S_n - \frac{(\Delta S_n)^2}{\Delta^2 S_n} = S_n -
\frac{(S_{n+1}-S_n)^2}{S_{n+2}-2 S_{n+1} + S_n}, \qquad n \in \setNz.
\end{equation}

$\Delta$ is the {\em forward difference operator}, which is defined as
\cite{Weniger,Summation,Concrete}
\begin{equation}
\Delta f(n) = f(n+1) - f(n).
\end{equation}
Powers of $\Delta$ are to be understood as multiple application of $\Delta$,
e.g. 
\begin{eqnarray}
\Delta^2 f(n) &=& \Delta \left[ \Delta f(n) \right] = \Delta \left[ f(n+1) - f(n)
\right] \nonumber \\
&=& f(n+2) - 2 f(n+1) + f(n).
\end{eqnarray}

Another method which can be found in the literature on numerical methods is
{\em Wynn's epsilon algorithm}, which is defined by the recursion scheme
\begin{eqnarray}
\epsilon_{-1}^{(n)} &=& 0 \nonumber \\
\epsilon_0^{(n)} &=& S_n \\
\epsilon_{k+1}^{(n)} &=& \epsilon_{k-1}^{(n+1)} + \frac{1}{\epsilon_k^{(n+1)} -
\epsilon_k^{(n)}}, \qquad k,n \in \setNz .\nonumber
\end{eqnarray}

We follow the notation of Ref. \cite{Weniger}.
The lower index indicates the transformation order, the upper index labels the
sequence elements.
The difference operator acts only on the upper index.

Similar to the epsilon algorithm is {\em Wynn's rho algorithm}:\nopagebreak
\begin{eqnarray}
\rho_{-1}^{(n)} &=& 0 \nonumber\\
\rho_0^{(n)} &=& S_n\\
\rho_{k+1}^{(n)} &=& \rho_{k-1}^{(n+1)} +
\frac{x_{n+k+1}-x_n}{\rho_k^{(n+1)}-\rho_k^{(n)}}, \qquad k,n \in \setNz,
\nonumber
\end{eqnarray}
where the {\em interpolation points} $x_k$ have to fulfill
$0 < x_i < x_j$ for $i<j$
and the sequence $\{x_n\}$ diverges to infinity.
A possible choice is $x_n=n+1$.

An example, where the rho algorithm is very powerful, is the series defining
$\zeta(2) = \sum_{i=1}^\infty \frac{1}{i^2}$. When one sums the series up to the
order $i=2000$ one gets only 4 correct decimal digits. Application of the rho
algorithm delivers more than 250 correct digits.

For the {\em Levin transformations} remainder
estimations have to be taken into account. The Levin transformation ist exact
for sequences of the form
\begin{equation}
S_n = S + \omega_n \,\sum_{j=0}^{k-1} \frac{c_j}{(n+\beta)^j}, \qquad k,n \in
\setNz,
\end{equation}
where $S$ is the limit of the sequence and $\beta > 0$ is a parameter which can,
in principle, be chosen arbitrarily and is usually set equal to 1. 
$\omega_n$ are the remainder estimates. They are, of course, generally not known
exactly, because the actual limit is unknown.
Therefore one uses assumptions for the $\omega_n$ which are customized to the given problem
and which define different kinds of Levin transformations.

As an example, the ansatz
\begin{equation}
\omega_n = \frac{a_n \, a_{n+1}}{a_n-a_{n+1}}, n \in \setNz,
\end{equation}
yields {\em Levin's v transformation}
\begin{equation}
v_k^{(n)}(\beta,S_n)=
\frac{
\sum\limits_{j=0}^k (-1)^j \, \binom{k}{j} \,
\frac{(\beta+n+j)^{k-1}}{(\beta+n+k)^{k-1}}\,
\frac{a_{n+j}-a_{n+j+1}}{a_{n+j}\, a_{n+j+1}} \,S_{n+j}
}{
\sum\limits_{j=0}^k (-1)^j\, \binom{k}{j} \,
\frac{(\beta+n+j)^{k-1}}{(\beta+n+k)^{k-1}}\,
\frac{a_{n+j}-a_{n+j+1}}{a_{n+j}\, a_{n+j+1}}
}.
\end{equation}

An enhancement of the epsilon algorithm is
{\em Brezinski's $theta$ algorithm}:\nopagebreak
\begin{eqnarray}
\vartheta_{-1}^{(n)} &=& 0 \nonumber\\
\vartheta_{0}^{(n)} &=& S_n \nonumber \\
\vartheta_{2k+1}^{(n)} &=& \vartheta_{2k-1}^{(n+1)} + \frac{1}{\Delta
\vartheta_{2k}^{(n)}} \\
\vartheta_{2k+2}^{(n)} &=& \vartheta_{2k}^{(n+1)} + \frac{\left[ \Delta
\vartheta_{2k}^{(n+1)} \right] 
\left[ \Delta \vartheta_{2k+1}^{(n+1)} \right]}{\Delta^2 \vartheta_{2k+1}^{(n)}},
\qquad k,n \in \setNz.
\nonumber
\end{eqnarray}

It is now suggestive to apply such a procedure, that improves the convergence of
a series, succesively. In Ref. \cite{Weniger} algorithms are discussed which
allow the calculation of high transformation orders efficiently and with low memory costs.

The principle is to calculate for each new element of the series all possible elements of the
matrix $\vartheta_i^j$. So one always gets the highest possible transformation
order. Details can be found in Ref. \cite{Weniger}, where also many more acceleration
methods are discussed.

For this work the algorithms mentioned above were programmed in \Fortran.
To achieve the desired numerical precision we used Bailey's library \MPFUN{} 
\cite{Bailey:mpfun,Bailey:transmp},
which allows calculations with arbitrary precision.


\section{Numerical Result for the Nonplanar Topology}
\label{nonplanar}
In this Section we sketch the application of the described methods to the
nonplanar diagram $N$ and show how a numerical result is obtained. 

The sum we have to calculate reads (\ref{NS})

\begin{equation}
\tilde{N}_S = (-4\eps-4\eps^2+32\eps^3+\mathcal{O}(\eps^4)) N_S ,
\end{equation}

\begin{equation}
N_S=\sum_{l=0}^\infty \sum_{m=0}^\infty \GSum{l}{m}{n}{g} T(\eps,l,m,n,g),
\end{equation}

with the addend terms
\begin{multline}
T(\eps,l,m,n,g) =
\frac{(1-\varepsilon)^4}{\Gamma(2-2\varepsilon)\,\Gamma^2(1-\varepsilon)}\\
\times \frac{1}{(l+1-\varepsilon)\,(m+1-\varepsilon)\,(n+1-\varepsilon)}\,
\frac{\Gamma(g+2-2\varepsilon)}{\Gamma(g+2-\varepsilon)}\\
\times \frac{\Gamma(g-l+1-\varepsilon)\,\Gamma(g-m+1-\varepsilon)\,\Gamma(g-n+1-\varepsilon)}
{\Gamma(g-l+1)\, \Gamma(g-m+1)\, \Gamma(g-n+1)}\cdot R(\eps,l,m,n),
\end{multline}
where
\begin{multline}
R(\eps,l,m,n) = \\
\frac{4}{\left( 2 + l + m + n - 6\,\varepsilon  \right) \,\left( n - 4\,\varepsilon  \right) \,
     \left( 1 + l + n - 4\,\varepsilon  \right) \,\left( 2 + l + m + n -
     4\,\varepsilon  \right) } \\
     + 
  \frac{4}{\left( 1 + l \right) \,\left( 4 + l + m + n \right) \,\left( 2 + l + m + n - 6\,\varepsilon  \right) \,
     \left( 1 + m - 2\,\varepsilon  \right) } + \ldots
\end{multline}
consists of rational expressions resulting from the radial
integration of Eq. (\ref{Radialintegral}).

\subsection{Coefficient of $\eps^1$}
The coefficient of $\eps^1$ is a triple sum of rational expressions, where no harmonic sums occur. The sum
can be evaluated analytically, which we
sketch briefly in this subsection.
The sum over $l$ can be performed easily and we are left with a double sum. In the summands we get
harmonic sums with arguments $m$ and $m+n$.

The resulting rational terms and terms with $S(m)$ can now be summed over $n$. The terms containing
$S(m+n)$ can be cast into an expression that can be summed over $m$ by means of a transformation
of variables. After the summation we are left with a simple sum with increased
number of terms. The summands now consist of rational expressions, harmonic sums
with simple argument $S_k(i+ n)$,
such with double argument $S_k(i+2n)$ and products of the latter. 

Again, the rational expressions can be summed up analytically by means of \Mathematica\ or
\SUMMER{}. Expressions containing harmonic sums with simple arguments can be
summed with \SUMMER{}.

Expressions with harmonic sums with double argument can be transformed to terms calculable with
\SUMMER\ by means of formula (\ref{S2x}).

A bit more complicated are terms where $2n$ appears in the denominator, like
\begin{equation*}
\sum_{n=1}^\infty \frac{S_2(n)}{(2n+1)^2}.
\end{equation*}

By means of Eq. (\ref{S(n)->S(2n)}) the harmonic sum is converted to sums with double argument
\begin{equation}
\sum_{n=1}^\infty \frac{2}{(2n+1)^2} \left[ S_{-2}(2n) + S_2(2n) \right].
\end{equation}
Then the sum is rewritten according to Eq. (\ref{sumf(2x)})  
\begin{equation}
\sum_{n=1}^\infty \frac{1+(-1)^n}{(n+1)^2}  \left[ S_{-2}(n) + S_2(n) \right] 
\end{equation}
and can now be calculated.
 
Finally, there are the products 
\begin{equation}
S_k(n+i) \, S_l(2m+j)
\end{equation}
which can be evaluated with the help of Eq. (\ref{S(n)->S(2n)}).

The problem for the $\eps^1$ coefficient can thus be solved completely and the
result reads
\begin{equation}
\tilde{N}_S|_{\eps^1} = 68\, \zeta_3^2 - 80\, \zeta_5 + 50\, \zeta_6,
\end{equation}
which agrees with \cite{Kazakov:MOU}.

\subsection{Coefficient of $\eps^2$}
In the expression for the $\eps^2$ coefficient the terms in the triple sum contain the harmonic sums 
$S_1(n-l)$, $S_1(m)$, $S_1(l)$ and $S_1(m+n)$. All expressions can be summed over a variable which does not
appear as an argument in $S$.

The result is then a double sum 
\begin{equation}
\sum_{m=1}^\infty \sum_{n=1}^\infty F(m,n),
\end{equation}
where terms with harmonic sums of the following types occur:
\begin{equation*}
\begin{array}{lclcl}
S_i(m),   & \quad & S_i(m)\,S_j(m),     & \quad & S_i(m)\,S_j(n), \\
S_i(m+n), & \quad & S_i(m+n)\,S_j(m+n), & \quad & S_i(m)\,S_j(m+n).\\
\end{array}
\end{equation*}

In the case of $S_i(m)$ and $S_i(m)\,S_j(m)$ we sum now over the variable $n$ which is not the argument of $S$.
Expressions containing $S_i(m+n)$, $S_i(m+n)\,S_j(m+n)$ can be handled by means of variable transformations as described in
Section \ref{nestedsums}.
The terms with $S_i(m)\,S_j(n)$, $S_i(m)\,S_j(m+n)$ can in the most cases be cast into a form to which algorithm B
of Ref. \cite{Uwer} can be applied. 

Thus the second summation yields a simple sum. Now the addends contain simple harmonic sums and
products of two or three harmonic sums, some of which have double arguments:
\begin{equation*}
S_i(m), \quad S_i(m)\,S_j(m), \quad S_i(m)\,S_j(2m), \quad
S_i(m)\,S_j(m)\,S_k(2m).
\end{equation*}
These terms can in principle be calculated by means of the described algorithms.

However, in some of the terms there are certain combinations of denominators and harmonic sums in which the
summation variable is doubled. These can in some cases be be handled with the known algorithms,
however, not in the general case. Therefore the problem of the order $\eps^2$ cannot
be solved analytically at present. To achieve this, more examinations in
the subject of the summations are necessary.  

\subsection{Numerical evaluation}
Since up to now an analytical solution is not possible with our method we try to find a numerical
solution. To do this, we simplify the terms as far as possible and calculate the
sums numerically.

For the numerical calculation the terms have been simplified with \Mathematica\ and output in
\Fortran\ syntax. This code has been optimized by means of \Maple\ and built in a \Fortran\ program.
To speed up the calculation this program was parallelized by means of \MPI{} \cite{MPI,MPI:Gropp}. Finally, the program was translated with Bailey's program \type{TRANSMP}
\cite{Bailey:transmp} to allow arbitrary numerical precision by means of the
\MPFUN{} library 
\cite{Bailey:mpfun}.
The programs store all partial sums. To the sequence of partial sums we apply
different convergence acceleration methods. 

A different approach to a numerical solution, is the following:
We begin with the representation of the sum (\ref{FormSumme})
\begin{equation}
\sum_{n=1}^\infty \sum_{m=0}^\infty \sum_{l=0}^{n} F(l+m,n+m-l,n).
\end{equation}
For a fixed $n=\nu$ the expression
\begin{equation}
\sum_{l=0}^{\nu} \sum_{m=0}^\infty  F(l+m,\nu+m-l,\nu)
\end{equation}
describes $\nu+1$ infinite sums
\begin{equation}
S_{\nu,l}= \sum_{m=0}^\infty  F(l+m,\nu+m-l,\nu).
\end{equation}

Sums of that kind can easily be calculated with \type{SUMMER}. So the obvious
procedure is to evaluate all contributions $S_{n,l}$ for all values of $n$ up to
an upper limit $n_{max}$.

Thus we get a sequence of expressions consisting of rational numbers and zeta
functions. These can be evaluated numerically to achieve a sequence of decimal
numbers $\{S_n\}$, where
\begin{equation*}
\lim_{n \to \infty} S_n = \tilde{N}_S|_{\eps^2}.
\end{equation*}
To this we again apply the convergence acceleration algorithms. 
The first elements of this sequence are
\begin{eqnarray*}
S_0 &=& 56\,{{\zeta }_2} - 56\,{{\zeta }_3} + 32\,{{\zeta }_2}\,{{\zeta }_3} - 
  80\,{{{\zeta }_3^2}} - 100\,{{\zeta }_4} + 204\,{{\zeta }_3}\,{{\zeta }_4} \\
  && + 64\,{{\zeta }_5} + 144\,{{\zeta }_2}\,{{\zeta }_5} - 488\,{{\zeta }_6} + 
  308\,{{\zeta }_7} ,\\
S_1 &=&  - \frac{1220}{3}\,{{\zeta }_2} - \frac{848}{9}\,{{\zeta }_3} + 
  160\,{{\zeta }_2}\,{{\zeta }_3} - 80\,{{{\zeta }_3^2}} - 
  \frac{644}{3}\,{{\zeta }_4} + 204\,{{\zeta }_3}\,{{\zeta }_4} \\
  && + 320\,{{\zeta }_5} + 144\,{{\zeta }_2}\,{{\zeta }_5} - 488\,{{\zeta }_6} + 
  308\,{{\zeta }_7} + \frac{3232}{9} ,\\
S_2 &=&   - \frac{35827}{108}\,{{\zeta }_2} - 
  \frac{2219}{18}\,{{\zeta }_3} + 168\,{{\zeta }_2}\,{{\zeta }_3} - 
  80\,{{{\zeta }_3^2}} - \frac{653}{2}\,{{\zeta }_4} + 
  204\,{{\zeta }_3}\,{{\zeta }_4} \\
  && + 396\,{{\zeta }_5} + 
  144\,{{\zeta }_2}\,{{\zeta }_5} - 488\,{{\zeta }_6} + 308\,{{\zeta }_7} + \frac{64279}{216}
\end{eqnarray*}
and the 500th element is (with rounded coefficients)
\begin{eqnarray*}
S_{500} &=& 109.191 - 65.7377\,{{\zeta }_2} - 114.167\,{{\zeta }_3} + 
  210.296\,{{\zeta }_2}\,{{\zeta }_3} - 80.\,{{{\zeta }_3^2}} - 
  741.581\,{{\zeta }_4} \\
 && + 204.\,{{\zeta }_3}\,{{\zeta }_4} + 
  500.591\,{{\zeta }_5} + 144.\,{{\zeta }_2}\,{{\zeta }_5} - 488.\,{{\zeta }_6} + 
  308.\,{{\zeta }_7}\\
  &\cong& 205.62576485415736523.
\end{eqnarray*}

This procedure can be implemented easily in \FORM{} or its  parallelized
version \PARFORM{} \cite{Fliegner,Tentyukov}.

We show in Table \ref{Tab_e2_2} the results of the \Fortran-Program. 
The entry in the line labeled `Sum' is the result we get by simply summing up
the terms, where the upper limit of the outermost summation was set to 15,000. The other 
lines contain the results obtained by application of
different algorithms.

\begin{table}[h]
\begin{center}
\begin{tabular}{|l|r|}
\hline
Method & \hfill Value for $\tilde{N}_S|_{\eps^2}$ \hfill{} \\ 
\hline
Sum      & 205.625 765 027 123 610 885 997 875 \\
\hline
Theta    & 205.625 765 027 124 189 154 823 016 \\
\hline
Epsilon  & 205.625 765 027 124 195 743 663 390 \\
\hline
Aitken   & 205.625 765 027 124 195 744 291 025 \\
\hline
Rho      & 205.625 765 027 124 195 744 492 088 \\
\hline
Levin v  & 205.625 765 027 124 196 134 222 058 \\
\hline
\end{tabular}
\caption{Numerical results for the $\eps^2$ coefficient of $N$ \label{Tab_e2_2}}
\end{center}
\end{table}
One can see, that the results of the Aitken and the Rho method agree in 21 decimal
digits. These algorithms delivered the best result 
\footnote{Summing up 2000 terms we get 3 correct digits for $P$. Application of the rho
algorithm yields 13 correct digits in this case.}
for the numerical
calculations of the planar topology, where the exact result was known and could
be used to check the results. Since the structure of the terms is similar, one
can expect, that they are also suitable for the nonplanar diagram. 
The result of the Epsilon procedure agrees in 20 digits with the former.
All results agree in 16 digits. In a very pessimistic interpretation at least these
digits can be considered reliable. 

The results of the numerical evaluation by means of the \FORM-method agree with these numbers in 14 or 20
digits, depending on the acceleration method used.

\subsection{Integer Relation Detection}
Finally, we want to get  an idea of the analytical result by
means of ``experimental mathematics''. 
Looking at the known result for the planar diagram, we assume, that the result
for $N$ be also a linear combination of zeta values with integer coefficients,
which have to be determined.
This problem is known as {\em integer relation detection} \cite{Bailey:IRD}.

In general a vector $\vek{x}=(x_1,x_2,...,x_n)$ with real or complex elements 
is said to possess an {\em integer relation}, if integers $a_i$ exist such that
at least one of them is different from zero and 
\begin{equation}
a_1 x_1 + a_2 x_2 + \ldots + a_n x_n = 0
\end{equation}

An algorithm which can solve this problem is the {\em \type{PSLQ} algorithm} 
\cite{PSLQ-Analysis,PSLQ-Programm,Bailey-Broadhurst}.

The input for this algorithm are in our case the numerically calculated number
and numerical values for the constants, i.e. the zeta values. Then the program
checks, whether there is an integer relation between the numbers within the
desired accuracy. We used an implementation of this method by O. Veretin
\cite{Veretin:pers}.

Assuming that the transcendentals are the same as for the planar diagram, as it is the
case in the lower orders in $\eps$, we find the following result for the 2nd
order coefficient of the $\eps$ expansion of $N$:
\begin{equation}
\tilde{N}_S|_{\eps^2} = -272\, \zeta_3^2 + 204\, \zeta_3\, \zeta_4 + 80 \zeta_5 - 200 \zeta_6 + 450
\zeta_7.
\end{equation}
This agrees with the analytical result which was found in the meantime
\cite{Baikov:unpublished,Chetyrkin:Talk}.

Comparison of this result with Table \ref{Tab_e2_2} shows, that all the digits
printed for the rho algorithm are correct. This proves the power of this
algorithm for our problem and justifies the confidence in the convergence
acceleration techniques retrospectively.


\section{Analytical Result for the Planar Topology}
\label{planar}
In this Section we show that the coefficient of $\eps^2$ for the planar diagram 
can be found analytically by means of the described method.
The steps we take are the same as described in Section \ref{nonplanar}. The
structure of the sums and of the terms are similar and so is the procedure of
the calculation. The difference is, the terms that cause problems in the case of
$N$, because we do not know algorithms for their evaluation, do not appear in
the case of $P$. 

After application of the algorithms we have the sum in the form
\begin{equation}
\label{P_Summe}
\tilde{P}_S|_{\eps^2} = P_0 + P_1 + \sum_{a=1}^\infty P_2(a),
\end{equation}
where $P_0$ is the contribution for $n=0$,
\begin{multline}
P_0 = 56\,\zeta_3 -164\,\zeta_4 + 56\,\zeta_5 -440\,\zeta_6 + 320\,\zeta_7 \\
+48\,\zeta_2\,\zeta_3 + 180\,\zeta_3\,\zeta_4 + 120\,\zeta_2\,\zeta_5 -
80\,\zeta_3^2,
\end{multline}
and
\begin{multline*}
P_1=
\frac{17686997}{17496} - \frac{166771}{324}\,\zeta_2 + \frac{86459}{162}\,\zeta_3 - 192\,\zeta_1\,\zeta_3 + 
  1152\,\zeta_2\,\zeta_3 \\
  - 16\,\zeta_1\,\zeta_2\,\zeta_3 - 592\,{\zeta_3^2} + 128\,\zeta_1\,{\zeta_3^2} - 
  \frac{2702}{9}\,\zeta_4 + 552\,\zeta_1\,\zeta_4 \\
  - 1032\,\zeta_3\,\zeta_4 - \frac{394}{3}\,\zeta_5 - 
  656\,\zeta_1\,\zeta_5 - 1240\,\zeta_2\,\zeta_5 - \frac{6770}{3}\,\zeta_6 \\
  + 770\,\zeta_1\,\zeta_6 + 2248\,\zeta_7
,
\end{multline*}
\begin{multline*}
P_2(a) = \frac{528\,S_1(a)}{a^5} - \frac{480\,S_1(a)}{a^4} -
\frac{48\,{S_1(a)}^2}{a^4} - \frac{48\,S_1(a)\,S_2(a)}{a^3} + \frac{6\,S_{1,3}(a)}{3 + a}\\
+ \frac{8\,{S_1(a)}^2\,S_2(a)}{{\left( 3 + a \right) }^3} +
\frac{32\,S_1(a)\,S_2(a)\,S_3(a)}{1 + a}  - 
  \frac{24\,S_1(a)\,S_{1,3}(a)}{{\left( 3 + a \right) }^2} + \ldots ,
\end{multline*}

$P_2(a)$ consists only of terms that can be calculated with \SUMMER. Doing this
we arrive at the known result
\begin{equation}
\tilde{P}_S|_{\eps^2} = -176\,{\zeta_3^2} + 132\,\zeta_3\,\zeta_4 +
80\,\zeta_5 - 200\,\zeta_6 + 317\,\zeta_7 .
\end{equation}


\section{Conclusion}
We described a method for the calculation of massless, dimensionally re\-gu\-la\-rized
Feynman integrals in their $\eps$-expansion. It is based on the expansion of the
integrand in terms of Gegenbauer polynomials. This yields sums over Gamma
functions which can be expanded in harmonic sums. As a consequence one arrives
at sums over
rational functions and products of such with harmonic sums which can be
simplified by the introduced summation algorithms and then be calculated
numerically or analytically.

The procedure in principle is independent of the Gegenbauer polynomial
technique. It can be useful wherever
there are sums of the described form.
The method should be expandable to more complicated problems, e.g. higher orders
of the diagrams under consideration or more complex diagrams. 

In the case of the nonplanar diagram we could not get to a complete analytical solution. To
achieve this, further knowledge in the summation techniques is required.

For this work we combined the abilities of different computer programs, in
particular the pattern matching facilities of \Mathematica, the code generation
functions of \Maple, and \FORM's power of managing very large expressions. 
It is the idea to implement the method in one programming language to make it
stand-alone. This could either be done in \FORM{} or in \type{C++}. The summation
algorithms of Ref. \cite{Uwer}  have already been implemented in both, the
\FORM{} version has recently been published \cite{xsummer}. In \cite{hypexp} also a
\Mathematica{} implementation was introduced.


\section{Summary of the results}
The coefficients of the $\eps$ expansion of the nonplanar three-loop-diagram $N$ could be 
calculated numerically to 20 digits 
 \begin{multline*}
\tilde{N}_S = 20.738555102867398527 + 66.168906981239990785 \,\eps \\
+ 205.62576502712419574 \,\eps^2
\end{multline*}

The analytical result for the $\order(\eps)$ term given in \cite{Kazakov:MOU}
could be reproduced,
\begin{equation}
\tilde{N}_S = 20\, \zeta_5 + \left( 68\, \zeta_3^2 - 80\, \zeta_5 + 50\,
\zeta_6 \right) \eps.
\end{equation}

Application of the \type{PSLQ} algorithm to the $\order(\eps^2)$ term gives
\begin{equation}
\tilde{N}_S|_{\eps^2} = -272\, \zeta_3^2 + 204\, \zeta_3\, \zeta_4 + 80 \zeta_5 - 200 \zeta_6 + 450
\zeta_7,
\end{equation}
which agrees with \cite{Baikov:unpublished,Chetyrkin:Talk}.

Also the analytical result up to order $\eps^2$ for the planar diagram
was verified.
\begin{equation}
\tilde{P}_S|_{\eps^2} = -176\,{\zeta_3^2} + 132\,\zeta_3\,\zeta_4 +
80\,\zeta_5 - 200\,\zeta_6 + 317\,\zeta_7 .
\end{equation}


\section*{Acknowledgements}
The author wishes to thank K. Chetyrkin for the suggestion to do
this work, for his support and information about GPXT and multiloop
calculations.

I thank P. Uwer, O. Veretin and S. Weinzierl for useful discussion about the harmonic sums and
related topics. I thank O. Veretin for providing me with his implementation of
the PSLQ algorithm.

E. J. Weniger from the University of Regensburg provided me with extensive
information about convergence acceleration methods. I am grateful for inspiring
correspondence in this subject.

Finally the author wishes to thank M. Steinhauser and K. Chetyrkin for reading the manuscript and many
useful comments.

This work was supported by the Graduiertenkolleg ``Hochenergiephysik und
Teilchenastrophysik'' and by the Sonderforschungsbereich TR 9, ``Computational
Particle Physics''.


\appendix

\section{Properties of the Gegenbauer Polynomials}
\label{Anhang_GP}
The {\em Gegenbauer polynomials}, also called {\em ultraspherical polynomials}, are the coefficients in the power series
exansion of the generating function
\begin{equation}
\frac{1}{(1-2rt+r^2)^\lambda} = \sum_{n=0}^\infty C_n^\lambda(t)\, r^n.
\end{equation}

They are orthogonal on the interval $[-1,+1]$ with the weight function 
\begin{equation}
w(x)=(1-x^2)^{(\lambda-\frac{1}{2})},
\end{equation}
\begin{equation}
\int_{-1}^{+1} C_m^\lambda(x)\, C_n^\lambda(x)\, (1-x^2)^{\lambda-\frac{1}{2}}
\,dx = \frac{\pi\,\Gamma(2\lambda+n)}{2^{2\lambda-1}
(\lambda+n)\,n!\,\Gamma^2(\lambda)} \,\delta_{mn}.
\label{GGP_OR}
\end{equation}

Special cases are the Chebyshew polynomials of first or second kind with
$\lambda=0, 1$, respectively, and the Legendre polynomials with
$\lambda=\frac{1}{2}$.

Information about these polynomials can be found in the books by Erd\'{e}lyi et
al. \cite{HTF},
Vilenkin \cite{Vilenkin} or Szeg\"o \cite{Szego}. 
The notation in this article follows Chetyrkin et al.
\cite{Chetyrkin}.

From the orthogonality relation (\ref{GGP_OR}) follows a relation for unit
vectors $\uvek{x},\uvek{y},\uvek{z}$,
which is used to reduce angular integrals in Section \ref{GPXT}:
\cite{Chetyrkin,HTF}
\begin{equation}
\label{GGP_ORS}
\int d\uvek{y}\, C_n^\lambda(\uvek{x} \cdot \uvek{y})\, C_m^\lambda(\uvek{y} \cdot
\uvek{z}) = \frac{\lambda}{n+\lambda}\, \delta_{n,m}\, C_n^\lambda(\uvek{x} \cdot
\uvek{z})
\end{equation}

Values for special parameters are:
\begin{equation}
\label{C0=1}
C_0^\lambda(x) = 1,
\end{equation}
\begin{equation}
\label{Cn1}
C_n^\lambda(1) = \frac{\Gamma(n+2\lambda)}{n!\,\Gamma(2\lambda)},
\end{equation}
\begin{equation}
C_n^\lambda(0) = \left\{ 
    \begin{array}{cc}
    0 & \quad n\, \text{odd} \\
    \frac{(-1)^m\, \Gamma(\lambda+m)}{m!\, \Gamma(\lambda)}, m=\frac{n}{2} &
    \quad n\,
    \text{even}
    \end{array}  \right. .
\end{equation}

Another basic feature of the Gegenbauer polynomials used in this work is the
expansion of a propagator:

\begin{equation}
\label{GGP-Entwicklung}
\frac{1}{(\vek{x}_1-\vek{x}_2)^{2\lambda}} = \frac{1}{\max(r_1,r_2)^\lambda} \,
\sum_{n=0}^\infty C_n^\lambda(\uvek{x}_1 \cdot \uvek{x}_2)\, \left\langle \frac{r_1}{r_2}
\right\rangle^{n/2}
\end{equation}
where
\begin{equation}
r_i = \vek{x}_i^2, \quad \uvek{x}=\frac{\vek{x}}{\sqrt{r}},
\end{equation}
\begin{equation}
\left\langle \frac{r_1}{r_2} \right\rangle
= \min \left( \frac{r_1}{r_2}\,, \frac{r_2}{r_1} \right)
= \left\{
   \begin{array}{cc}
   r_1 / r_2 & \text{if}\,r_1 \leq r_2 \\
   r_2 / r_1 & \text{if}\,r_2 \leq r_1 
   \end{array}  \right. .
\end{equation}

The exponential function can be expanded by means of
\begin{equation}
\label{Exp-Entwicklung}
e^{2i\vek{p}\vek{x}} = \Gamma(\lambda) \sum_{n=0}^\infty i^n\, (n+\lambda)\,
C_n^\lambda (\uvek{x} \cdot \uvek{p})\, (p^2r)^{n/2}\, j_{\lambda+n}(p^2r),
\end{equation}
where $j_\alpha(z)$ is related to the  Bessel function $J_\alpha(z)$ by
\begin{equation}
J_\alpha(z) = (\sfrac{1}{2}z)^\alpha\, j_\alpha(\sfrac{1}{4}z^2),
\end{equation}
and the following equation holds:
\begin{equation}
\label{Besselintegral}
\int_0^\infty z^b\,j_\alpha(z)\,dz = \frac{\Gamma(b+1)}{\Gamma(a-b)}, \quad
\begin{array}{l} \Re b > -1 \\ \Re a > 2\, \Re b + \sfrac{1}{2}. \end{array}
\end{equation}

Finally, a formula is needed which allows to rewrite the product of two
Gegenbauer polynomials as sum of single Gegenbauer polynomials.
\begin{equation}
\label{CC-Summe}
C_l^\lambda(x)\,C_m^\lambda(x) = \sum_{\substack{n=|l-m| \\ l+m+n=2g,\, g \in
\mathds{Z}}}^{l+m} D_\lambda(l,m;n)\,C_n^\lambda(x),
\end{equation}
\begin{multline}
D_\lambda(l,m;n) =
\frac{n!\,(n+\lambda)\,\Gamma(g+2\lambda)}{\Gamma^2(\lambda)\,\Gamma(g+\lambda+1)\,\Gamma(n+2\lambda)}\,
\\
\times\,\frac{\Gamma(g-l+\lambda)\,\Gamma(g-m+\lambda)\,\Gamma(g-n+\lambda)}{\Gamma(g-l+1)\,\Gamma(g-m+1)\,\Gamma(g-n+1)}.
\end{multline}

\newcommand{\rauthor}[1]{#1}
\newcommand{\rtitle}[1]{\textit{#1}}
\newcommand{\rjour}[4]{\mbox{#1 \textbf{#2} (#3), #4}}
\newcommand{\ajour}[4]{#1 \textbf{#2} (#3), #4}
\newcommand{\rhepnum}[1]{#1}
\newcommand{\rnote}[1]{#1}
\newcommand{\rbook}[2]{\mbox{#1 #2}}
\newcommand{\rweb}[1]{\texttt{#1}}

\end{document}